\documentclass[aps,prb,floatfix,twocolumn,showpacs]{revtex4}%
\usepackage{amsmath}
\usepackage{graphicx}
\usepackage{amsfonts}
\usepackage{amssymb}%
\providecommand{\U}[1]{\protect\rule{.1in}{.1in}}
\begin{document}
\title{Multi-terminal spin-dependent transport in ballistic carbon nanotubes}
\author{Audrey Cottet, Ch\'{e}ryl Feuillet-Palma and Takis Kontos}
\affiliation{Ecole Normale Sup\'{e}rieure, Laboratoire Pierre Aigrain, 24 rue Lhomond,
75231 Paris Cedex 05, France}


\begin{abstract}
We study theoretically non-local spin-transport in a ballistic carbon nanotube
contacted to two ferromagnetic leads and two normal metal leads. When the
magnetizations of the two ferromagnets are changed from a parallel to an
antiparallel configuration, the circuit shows an hysteretic behavior which is
specific to the few-channels regime. In the coherent limit, the amplitude of
the magnetic signals is strongly enhanced due to resonance effects occuring
inside the nanotube. Our calculations pave the way to new experiments on
low-dimensional non-local spin-transport, which should give results remarkably
different from the experiments realized so far in the multichannel diffusive
incoherent regime.

\end{abstract}
\date{\today}
\maketitle

\section{Introduction}

Non-local electric effects have been observed since the early days of
mesoscopic physics, e.g. in metallic circuits\cite{Webb:89,FilsMetal}. This
fact is related to the primarily non-local nature of electronic wave functions
in quantum coherent conductors. The spin-degree of freedom has raised little
attention in this context, although its control and detection is one of the
major challenges of nanophysics, nowadays. Non-local spin signals have been
studied for multi-terminal metallic
conductors\cite{Johnson,Jedema,Jedema2,Zaffalon}, semiconductors\cite{Lou} and
graphene\cite{Tombros2}, in the multichannel diffusive incoherent (MDI)
regime. It has been found that a non-equilibrium spin accumulation induced by
a ferromagnet into a given conductor can be detected as a voltage across the
interface between this conductor and another ferromagnet\cite{Silsbee}.
However, to our knowledge, spin-dependent non local effects have not been
investigated in the coherent regime, so far.

Carbon-nanotubes-based circuits are appealing candidates for observing a
non-local, spin-dependent, and coherent behavior of electrons. First,
electronic transport in carbon nanotubes (CNTs) can reach the few-channels
ballistic regime, as suggested by the observation of Fabry-Perot-like
interference patterns\cite{Liang}. Secondly, spin injection has already been
demonstrated in CNTs connected to two ferromagnetic leads (see
Ref.~\onlinecite {SST} for a review). Thirdly, non local voltages have been
observed in CNTs contacted to four normal-metal leads\cite{Makarovski}, which
suggests that electrons can propagate in the nanotube sections below the
contacts. The study of non-local spin transport in CNTs has recently triggered
some experimental efforts\cite{Tombros,Gunnar,Cheryl}. However, a theoretical
insight on this topic is lacking. Some major questions to address are what are
the signatures of a non-local and spin-dependent behavior of electrons in a
nanoconductor, and to which extent these signatures are specific to the
coherent regime or the few-channels case. \begin{figure}[ptbh]
\includegraphics[width=1.\linewidth]{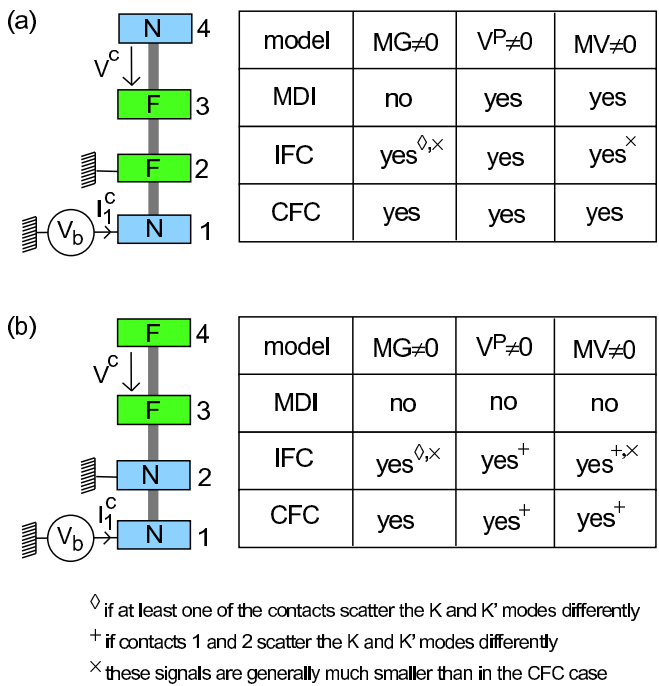}\caption{Left: The two types
of circuits [setups (a) and (b)] studied in this article. The central
conductor (represented with the grey bar) is contacted to two normal metal
leads $N$ and two ferromagnetic leads $F$, which can be magnetized in parallel
($c=P$) or antiparallel ($c=AP$) configurations. The only difference between
setups (a) and (b) is the position of the two $F$ leads. Contacts 1 and 2 are
used as source and drain to measure a local conductance $G^{c\text{ }}$and
contacts 3 and 4 are used to probe a non-local voltage $V^{c}$ outside the
classical current path. Right: Tables presenting the behaviors of setups (a)
and (b) in various regimes. We study the existence of the signals
$MG=(G^{P}-G^{AP})/G^{P}$, $V^{P}$ and $MV=(V^{P}-V^{AP})/V_{b}$. We compare
the predictions of the coherent four-channels model (CFC) of section
\ref{ScatModel}, the incoherent four-channels model (IFC) of section \ref{IFC}
and the multichannel diffusive incoherent (MDI) model of section \ref{MDI}.}%
\label{Circuits}%
\end{figure}

In this paper, we study the behavior of a CNT with two normal metal ($N$)
leads and two ferromagnetic ($F$) leads magnetized in colinear directions. Two
leads are used as source and drain to define a local conductance $G^{c\text{
}}$and the other two are used to probe a non-local voltage $V^{c}$ outside the
classical current path. We consider two different setups which differ on the
positions of the $F$ leads. Setup (a) corresponds to the standard geometry
used for the study of the MDI limit. In setup (b), the two $F$ leads play the
role of the voltage probes, so that no magnetic response is allowed in the MDI
limit. We mainly focus on the coherent regime, using a scattering description
with two transverse modes, to account for the twofold orbital degeneracy
commonly observed in CNTs\cite{deg}. This minimal description is appropriate
at low temperatures and bias voltages. We take into account both the
spin-polarization of the tunneling probabilities at the ferromagnetic contacts
and the Spin-Dependence of Interfacial Phase Shifts (SDIPS) which has been
shown to affect significantly spin-dependent transport in the two-terminals
case\cite{Cottet06a,Cottet06b,Sahoo}. This approach leads to strong
qualitative differences with the MDI case. In particular, we find a magnetic
signal in the conductance $G^{c}$ of setups (a) and (b), which would not occur
in the MDI limit. We also predict an unprecedented magnetic signal in $V^{c}$
for setup (b). We find that these effects already arise in the incoherent few
channels regime. However, they are much stronger in the coherent case, due to
resonances which occur inside the CNT. These resonances make the circuit
sensitive to the SDIPS, which can furthermore enhance the amplitude of the
magnetic signals.

This paper is organized as follows: section II defines setups (a) and (b),
section III discusses the multichannel diffusive incoherent (MDI) limit,
section IV focuses on the coherent\ four-channels (CFC) scattering
description, section V presents an incoherent four-channels (IFC) description,
section VI discusses the experimental results presently available, and section
VII concludes.

\section{Definition of setups (a) and (b)}

In this article, we consider a central conductor (CC) connected to an ensemble
$\mathcal{L}$ of two ferromagnetic ($F$) and two normal metal ($N$)
reservoirs. We study the two configurations presented in Fig.~\ref{Circuits}.
In both cases, lead 1 is connected to a bias voltage source $V_{b}$, lead 2 is
connected to ground, whereas leads 3 and 4 are left floating. The only
difference between setups (a) and (b) is the position of the two $F$ leads.
These $F$ leads can be magnetized in parallel ($c=P$) or antiparallel ($c=AP$)
configurations. We will study the conductance $G^{c}=\partial I_{1}%
^{c}/\partial V_{b}$ between contacts 1 and 2 and the voltage drop $V^{c}$
between leads 3 and 4. The dependence of these quantities on the magnetic
configuration $c$ of the ferromagnetic electrodes can be characterized with
the magnetic signals $MG=(G^{P}-G^{AP})/G^{P}$ and $MV=(V^{P}-V^{AP})/V_{b}$.

\section{Multichannel diffusive incoherent limit\label{MDI}}

We first briefly discuss the behavior of setups (a) and (b) in the
multichannel diffusive incoherent (MDI) regime. This case has been thoroughly
investigated, in relation with experiments in which the CC is a metallic
island\cite{Johnson,Jedema,Jedema2,Zaffalon}. For a theoretical description of
this regime, one can use spin-currents and a spin-dependent electrochemical
potential $\mu_{\sigma}$ which obey a local spin-dependent Ohm's law, provided
the mean free path in the sample is much shorter than the spin-flip length. We
refer the reader to Ref.~\onlinecite {Valet} for a detailed justification of
this approach from the Boltzmann equations, and to Ref.~\onlinecite {Zutic}
for an overview of this field of research. In this section, we summarize the
behaviors expected for setups (a) and (b) in the MDI limit (see Appendix A for
a short derivation of these results from a resistors model). A finite current
between leads 1 and 2 can lead to a spin accumulation (i.e. $\mu_{\uparrow
}\neq\mu_{\downarrow}$) in the CC if lead 1 or 2 is ferromagnetic, because
spins are injected into and extracted from the CC\ with different rates in
this case. The spin accumulation diffuses along the CC beyond lead 2, and
reaches leads 3 and 4, provided the spin-flip length is sufficiently long.
Then, leads $3$ and $4$ can be used to detect the spin accumulation provided
one of them is ferromagnetic. Indeed, a local unbalance $\mu_{\uparrow}\neq
\mu_{\downarrow}$ in the CC will produce a voltage drop between the floating
lead $j\in\{3,4\}$ and the CC if $j$ is ferromagnetic (this voltage drops aims
at equilibrating the spin currents between the CC and the ferromagnetic
contact). One can thus conclude that in setup (a), a spin accumulation occurs
when $V_{b}\neq0$, which leads to $V^{c}=V^{3}-V^{4}\neq0$. In contrast, one
finds $V^{c}=0$ in setup (b) because a current flow between the $N$ leads 1
and 2 cannot produce any spin accumulation. For completeness, we also mention
that in the MDI limit, one finds $G^{P}=G^{AP}$ for both setups (a) and (b),
due to the fact that leads 3 and 4 are left floating (see Appendix A). The
table in Fig. \ref{Circuits} summarizes these results.

\section{Coherent four-channels limit\label{ScatModel}}

\subsection{General scattering description}

\begin{figure}[ptbh]
\includegraphics[width=0.7\linewidth]{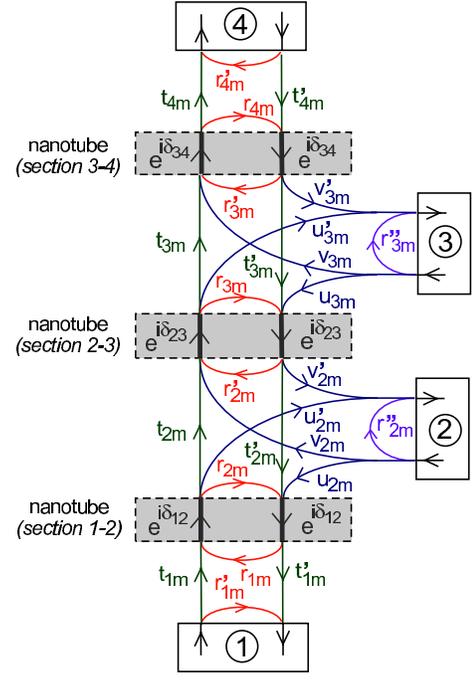}\caption{Scheme
representing the notations used for the scattering amplitudes of channel $m$
in setups (a) and (b). The rectangles with full lines represent the different
leads and the dashed rectangles represent the nanotube sections between those
leads. We note $\delta_{jk}$\ the phase shift acquired by electrons while
propagating along the nanotube between contacts $j$\ and $k$. At this stage,
we use $\delta_{jk}=\delta_{kj}$, but we keep $t_{jm}\neq t_{jm}^{\prime}$,
$u_{jm}\neq u_{jm}^{\prime}$, and $v_{jm}\neq v_{jm}^{\prime}$ for
transparency of the calculation.}%
\label{Amplitudes}%
\end{figure}

\begin{figure}[ptbh]
\includegraphics[width=0.65\linewidth]{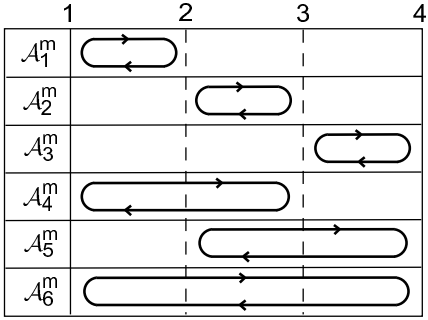}\caption{Scheme
representing different resonances (noted $\mathcal{A}_{i}^{m}$) which can
occur in setups (a) and (b) when $t_{2[3],m}=t_{2[3],m}^{^{\prime}}=0$ or $1$.
The upper numbers indicate the position of contacts 1, 2, 3 and 4.}%
\label{Reson4term}%
\end{figure}

In this section, we study the case where the CC is a ballistic carbon nanotube
(CNT) allowing coherent transport. The observation of Fabry-Perot like
interference patterns\cite{Liang} suggests that it is possible, with certain
type of metallic contacts, to neglect electronic interactions inside CNTs. We
thus use a Landauer-B\"{u}ttiker scattering description\cite{Buttiker1}. We
take into account two transverse modes $p\in\{K,K^{\prime}\}$, to account for
the twofold orbital degeneracy commonly observed in CNTs\cite{deg}. Each
transverse mode one has two spin submodes $\sigma\in\{\uparrow,\downarrow\}$,
defined colinearly to the polarization of the $F$ leads. This gives four
channels $m=(p,\sigma)$ in total. We assume that spin is conserved upon
scattering by the CNT/lead interfaces and upon propagation along the CNT. This
requires, in particular, that the magnetization direction can be considered as
uniform in the four $F$ leads, and that spin-orbit coupling and spin-flip
effects can be neglected inside the CNT and upon interfacial scattering. For
simplicity, we also assume that the transverse index $p$ is conserved. In the
linear regime, the average current through lead $j$ writes%

\begin{equation}
I_{j}^{c}=\sum_{k}G_{jk}V_{k}^{c} \label{I}%
\end{equation}
with
\begin{equation}
G_{jk}=G_{K}[4\delta_{jk}-\sum_{m}\left\vert S_{jk}^{m}\right\vert ^{2}]
\label{Gdef}%
\end{equation}
$G_{K}=e^{2}/h$, and $S_{jk}^{m}$ the scattering amplitude from lead $k$ to
lead $j$ for electrons of channel $m$. Equation (\ref{I}) involves the
electrostatic potential $V_{k}^{c}$ of lead $k$ (we assume that the leads are
in local equilibrium, so that each one has a single chemical potential for
both spin directions). Note that $G_{jk}$ and $S_{jk}^{m}$ implicitly depend
on the configuration $c$ of the ferromagnetic electrodes. In this section, we
calculate $G^{c}$ and $V^{c}$ by using the general notations of
Fig.~\ref{Amplitudes} for the scattering amplitudes. \ The phase shift $\delta_{jk}$ acquired by electrons along the
CNT from contacts $j$ to $k$ can be considered as independent from $m$, with
$\delta_{jk}=\delta_{kj}$\cite{Saito}. In practice, $\delta_{12}$,
$\delta_{23}$, and $\delta_{34}$ can be tuned using local gate voltage
electrodes to change the electronic wavevector in the different CNT sections
\cite{LocalGates}. We will thus study the signals $G^{c}$, $V^{c}$, $MG$ and
$MV$ as a function of these phases. The calculation of the voltage drop
$V^{c}$ requires to determine $V_{3}^{c}$ and $V_{4}^{c}$ from $\left\langle
I_{3}\right\rangle =\left\langle I_{4}\right\rangle =0$. This yields:
\begin{equation}
\frac{V^{c}}{V_{b}}=\frac{G_{41}G_{32}-G_{42}G_{31}}{G_{34}G_{43}-G_{33}%
G_{44}} \label{Vnl4pat}%
\end{equation}
and
\begin{align}
G^{c}/G_{K}  &  =G_{11}+[G_{13}(G_{44}G_{31}-G_{41}G_{34})\nonumber\\
&  +G_{14}(G_{41}G_{33}-G_{31}G_{43})]/[G_{34}G_{43}-G_{33}G_{44}]
\end{align}
Using the notations of Fig.~\ref{Amplitudes}, the elements $|S_{jk}^{m}|$
occurring in Eq.~(\ref{Vnl4pat}) through the coefficients $G_{jk}$ of Eq.
(\ref{Gdef}) can be calculated as
\begin{equation}
\left\vert S_{41}^{m}\right\vert =\left\vert D_{m}^{-1}t_{1m}t_{2m}%
t_{3m}t_{4m}\right\vert \label{S41quat}%
\end{equation}%
\begin{align}
\left\vert S_{31}^{m}\right\vert  &  =\left\vert D_{m}^{-1}t_{1m}%
t_{2m}\right\vert \nonumber\\
&  \times\left\vert u_{3m}^{^{\prime}}+r_{4m}(t_{3m}v_{3m}^{\prime}%
-r_{3m}^{\prime}u_{3m}^{\prime})e^{i2\delta_{34}}\right\vert \label{S31quat}%
\end{align}%
\begin{align}
\left\vert S_{42}^{m}\right\vert  &  =\left\vert D_{m}^{-1}t_{3m}%
t_{4m}\right\vert \nonumber\\
&  \times\left\vert v_{2m}+r_{1m}(t_{2m}u_{2m}-r_{2m}v_{2m})e^{i2\delta_{12}%
}\right\vert \label{S42quat}%
\end{align}%
\begin{equation}
S_{32}^{m}=S_{31}^{m}S_{42}^{m}/S_{41}^{m} \label{S32quat}%
\end{equation}%
\begin{align}
\left\vert S_{34}^{m}\right\vert  &  =\left\vert D_{m}^{-1}t_{4m}^{\prime
}\right\vert \times\left\vert v_{3m}^{\prime}(1-r_{1m}r_{2m}e^{i2\delta_{12}%
})\right. \nonumber\\
&  +e^{i2\delta_{23}}(r_{2m}^{\prime}+r_{1m}e^{i2\delta_{12}}\left[
t_{2m}t_{2m}^{\prime}-r_{2m}r_{2m}^{\prime}\right]  )\nonumber\\
&  \times\left.  (t_{3m}^{\prime}u_{3m}^{\prime}-r_{3m}v_{3m}^{\prime
})\right\vert \label{S34quat}%
\end{align}
and
\begin{align}
\left\vert S_{43}^{m}\right\vert  &  =\left\vert D_{m}^{-1}t_{4m}\right\vert
\times\left\vert v_{3m}[1-r_{1m}r_{2m}e^{i2\delta_{12}}]\right. \nonumber\\
&  +e^{i2\delta_{23}}[r_{2m}^{\prime}+r_{1m}e^{i2\delta_{12}}\left(
t_{2m}t_{2m}^{\prime}-r_{2m}r_{2m}^{\prime}\right)  ]\nonumber\\
&  \times\left.  \lbrack t_{3m}u_{3m}-r_{3m}v_{3m}]\right\vert \label{S43quat}%
\end{align}
with
\begin{align}
D_{m}  &  =[\left(  1-r_{1m}r_{2m}e^{i2\delta_{12}}\right)  \left(
1-r_{2m}^{\prime}r_{3m}e^{i2\delta_{23}}\right) \nonumber\\
&  \times\left(  1-r_{3m}^{^{\prime}}r_{4m}e^{i2\delta_{34}}\right)
]\nonumber\\
&  -t_{2m}t_{2m}^{\prime}r_{1m}r_{3m}\left(  1-r_{3m}^{^{\prime}}%
r_{4m}e^{i2\delta_{34}}\right)  e^{i2(\delta_{12}+\delta_{23})}\nonumber\\
&  -t_{3m}t_{3m}^{\prime}r_{2m}^{\prime}r_{4m}\left(  1-r_{1m}r_{2m}%
e^{i2\delta_{12}}\right)  e^{i2(\delta_{23}+\delta_{34})}\nonumber\\
&  -t_{2m}t_{2m}^{\prime}t_{3m}t_{3m}^{\prime}r_{1m}r_{4m}e^{i2(\delta
_{12}+\delta_{23}+\delta_{34})} \label{dquat}%
\end{align}
The missing coefficients $G_{33}$ and $G_{44}$ can be obtained from the above
Eqs. using $G_{33}=-(G_{34}+G_{31}+G_{32})$ and $G_{44}=-(G_{43}+G_{41}%
+G_{42})$. For calculating $G^{c}$, one furthermore needs
\begin{equation}
\left\vert S_{14}^{m}\right\vert =\left\vert D_{m}^{-1}t_{1m}^{\prime}%
t_{2m}^{\prime}t_{3m}^{\prime}t_{4m}^{\prime}\right\vert \label{S14quat}%
\end{equation}%
\begin{align}
\left\vert S_{13}^{m}\right\vert  &  =\left\vert D_{m}^{-1}t_{1m}^{\prime
}t_{2m}^{\prime}\right\vert \nonumber\\
&  \times\left\vert u_{3m}+r_{4m}e^{i2\delta_{34}}(t_{3m}^{\prime}%
v_{3m}-r_{3m}^{\prime}u_{3m})\right\vert
\end{align}
and
\begin{align}
S_{11}^{m}-r_{1m}^{\prime}  &  =D_{m}^{-1}t_{1m}t_{1m}^{\prime}e^{i2\delta
_{12}}\{r_{2m}\left(  1-r_{3m}^{\prime}r_{4m}e^{i2\delta_{34}}\right)
\nonumber\\
&  +e^{i2\delta_{23}}[r_{3m}+r_{4m}e^{i2\delta_{34}}(t_{3m}t_{3m}^{\prime
}-r_{3m}r_{3m}^{\prime})]\nonumber\\
&  \times\lbrack t_{2m}t_{2m}^{\prime}-r_{2m}r_{2m}^{\prime}]\}
\label{S11quat}%
\end{align}
The denominator $D_{m}$ accounts for multiple resonances inside the CNT.
Figure~\ref{Reson4term} depicts some resonances $\mathcal{A}_{n}^{m}$, with
$n\in\lbrack1,6]$, which can occur in limiting cases where $t_{2[3],m}%
=t_{2[3],m}^{^{\prime}}=0$ or $1$. In the general case, Eq. (\ref{dquat}) indicates that these
different resonances are coupled. For $t_{2m}=t_{2m}^{^{\prime}}=0$, $G^{c}$
corresponds to the conductance of a two-terminals device, independent from
$\delta_{23}$ and $\delta_{34}$, and $V^{c}$ vanishes. For $t_{2m}%
=t_{2m}^{^{\prime}}\neq0$ and $t_{3m}=t_{3m}^{^{\prime}}=0$, $G^{c}$ depends
on $\delta_{12}$ and $\delta_{23}$, but not on $\delta_{34}$, and $V^{c}$
still vanishes. Having a non-local signal $V^{c}\neq0$ requires a direct
CNT-CNT transmission at both contacts $2$ and $3$. It also requires that the
four channels $m$ are not coupled to the leads in the same way. Indeed, from
Eqs.~(\ref{Vnl4pat}) and (\ref{S32quat}), one can check that if all the
$S_{jk}^{m}$ coefficients are independent from $m$, one finds $V^{c}=0$ due to
the series structure of the device \cite{Baranger}. Interestingly, a finite
$V^{c}$ has already been obtained in a CNT connected to four normal-metal
leads\cite{Makarovski}, which suggests that the $K$ and $K^{\prime}$ modes
were not similarly coupled to those leads. In principle, such an asymmetry is
also possible with ferromagnetic contacts.

\subsection{Parametrization of the lead/nanotube contacts}

In the following, we assume that the top and bottom halves of the three
terminals contacts $j\in\{2,3\}$ in Fig.~\ref{Circuits} are symmetric. We
furthermore take into account that the scattering matrix associated to each
contact is invariant upon transposition, due to spin-conservation\cite{SC}.
This gives $t_{jm}=t_{jm}^{\prime}$, $r_{jm}=r_{jm}^{\prime}$, and
$u_{jm}=u_{jm}^{\prime}=v_{jm}=v_{jm}^{\prime}$ for $j\in\{2,3\}$. In this
case, on can check from Eqs. (\ref{S41quat}-\ref{S11quat}) that\ $G^{c}$ and
$V^{c}$ depend only on six interfacial scattering phases, i.e. those of
$r_{1m}$, $t_{2m}$, $r_{2m}$, $t_{3m}$, $r_{3m}$ and $r_{4m}$, which
correspond to processes during which electrons remain inside the CNT
\cite{simpl}. For contacts $j\in\{2,3\}$, it is thus convenient to use the
parametrization%
\begin{equation}
t_{j(p,\sigma)}=\sqrt{T_{j,p}(1+\sigma P_{j,p})}e^{i(\varphi_{j,p}^{T}%
+\frac{\sigma}{2}\Delta\varphi_{j,p}^{T})}%
\end{equation}%
\begin{align}
r_{j(p,\sigma)}  &  =\left[  \sqrt{1-\left\vert t_{j(p,\sigma)}\sin[\phi
_{j}^{(p,\sigma)}]\right\vert ^{2}}\right. \nonumber\\
&  +\left.  \left\vert t_{j(p,\sigma)}\right\vert \cos[\phi_{j}^{(p,\sigma
)}]\right]  e^{i(\varphi_{j,p}^{R}+\frac{\sigma}{2}\Delta\varphi_{j,p}^{R})}%
\end{align}%
\begin{equation}
\left\vert u_{jm}\right\vert =\sqrt{1-\left\vert r_{jm}\right\vert
^{2}-\left\vert t_{jm}\right\vert ^{2}} \label{K}%
\end{equation}
with
\[
\phi_{j}^{(p,\sigma)}=\varphi_{j,p}^{R}-\varphi_{j,p}^{T}+\frac{\sigma}%
{2}(\Delta\varphi_{j,p}^{R}-\Delta\varphi_{j,p}^{T})
\]
The above expressions depend on six real parameters $T_{j,p}$, $P_{j,p}$,
$\varphi_{j,p}^{T}$, $\varphi_{j,p}^{R}$, $\Delta\varphi_{j,p}^{R}$ and
$\Delta\varphi_{j,p}^{T}$\cite{param}. In order to have unitary lead/CNT
scattering matrices, on must use $0\leq T_{j,p}(1+\sigma P_{j,p})\leq1$
and\cite{annul} $\pi/2\leq\phi_{j}^{m}[2\pi]\leq3\pi/2$. These conditions
imply $0\leq|t_{jm}|^{2}\leq1$, $0\leq|r_{jm}|^{2}\leq1$ and\cite{sym}
$0\leq|u_{jm}|^{2}\leq1/2$. For contacts $j\in\{1,4\}$, one can use%
\begin{equation}
\left\vert t_{j}^{(p,\sigma)}\right\vert =\left\vert t_{j}^{^{\prime}%
(p,\sigma)}\right\vert =\sqrt{T_{j,p}(1+\sigma P_{j,p})}%
\end{equation}
and%
\begin{equation}
\arg(r_{j}^{(p,\sigma)})=c_{p}\varphi_{j}^{R}+\frac{\sigma}{2}\Delta
\varphi_{j,p}^{R} \label{phi1}%
\end{equation}
with $c_{K(K^{\prime})}=\pm1$. In Eq.~(\ref{phi1}), we have assumed
$\sum_{p,\sigma}\arg(r_{1}^{(p,\sigma)})=0$ and $\sum_{p,\sigma}\arg
(r_{4}^{(p,\sigma)})=0$, because, from Eqs. (\ref{S41quat}-\ref{S11quat}),
these quantities only shift the variations of $G^{c}$, $V^{c}$, $MG$ and $MV$
with respect to $\delta_{12}$ and $\delta_{34}$, respectively. The parameters
$P_{j,p}$, with $j\in\{1,2,3,4\}$, produce a spin-polarization of the
transmission probabilities $|t_{j}^{m}|^{2}$. The parameters $\Delta
\varphi_{j,p}^{R(T)}$ allow to take into account the Spin-Dependence of
Interfacial Phase Shifts (SDIPS), which has already been shown to affect
significantly the behavior of CNT spin-valves\cite{Cottet06a,Cottet06b,Sahoo}.
We will show below that the SDIPS also modifies the behavior of multi-terminal
setups. Note that for $j\in\{2,3\}$, the parameters $\Delta\varphi
_{p,j}^{R(T)}$ also contribute to the spin dependence of $|r_{jm}|^{2}$ and
$|u_{jm}|^{2}$: the SDIPS and the spin-dependence of interfacial scattering
probabilities are not independent in three-terminal contacts, due to the
unitarity of scattering processes.

\subsection{Behavior of setup (a)}

\begin{figure}[ptbh]
\includegraphics[width=1.\linewidth]{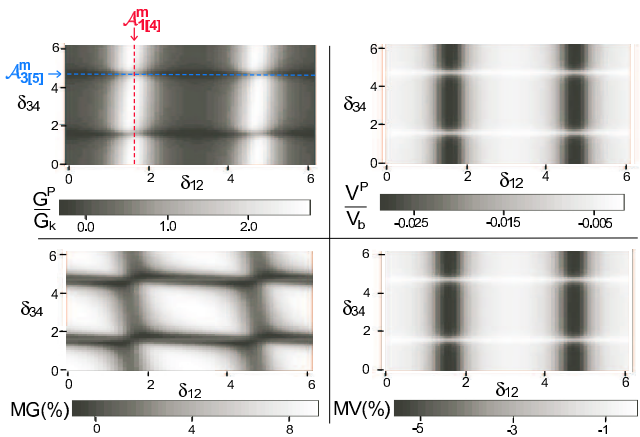}\caption{Signals
$G^{P}$ (top left panel), $V^{P}$ (top right panel), $MG$ (bottom left panel),
and $MV$ (bottom right panel) as a function of $\delta_{12}$ (horizontal axes)
and $\delta_{34}$ (vertical axes) for a setup (a) with symmetric $K$ and
$K^{\prime}$ channels. We have used $T_{1,K[K^{\prime}]}$ $=0.6$,
$T_{2,K[K^{\prime}]}$ $=0.1$, $T_{(3)4,K[K^{\prime}]}$ $=0.3$,
$P_{2[3],K[K^{\prime}]}$ $=0.4$, $\varphi_{1(4)}^{R}=0$, $\varphi
_{2(3),K[K^{\prime}]}^{R}=\pi$, $\varphi_{2[3],K[K^{\prime}]}^{T}=0$,
$\Delta\varphi_{2(3),K[K^{\prime}]}^{R\{T\}}=0$ and $\delta_{23}=\pi$. We have
indicated the position of the resonances $\mathcal{A}_{1(4)}^{m}$ and
$\mathcal{A}_{3(5)}^{m}$ with red and blue dashed lines, respectively.}%
\label{GraySetupAbis}%
\end{figure}
\begin{figure}[ptbh]
\includegraphics[width=1.\linewidth]{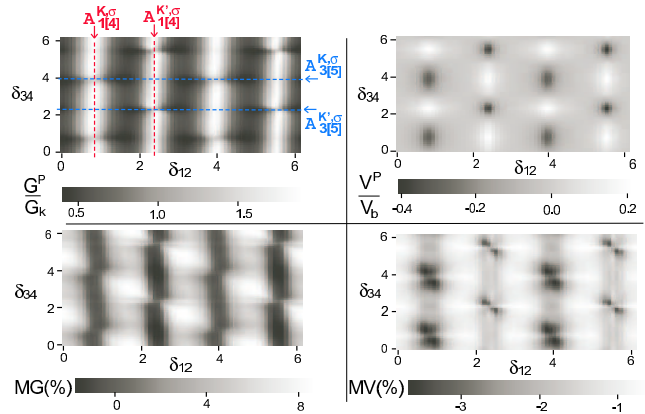}\caption{Signals $G^{P}$,
$V^{P}$ , $MG$, and $MV$ as a function of $\delta_{12}$ and $\delta_{34}$, for
a setup (a) with dissymmetric $K$ and $K^{\prime}$ channels. We have used
$T_{1[4],K}=0.5$, $T_{1[4],K^{\prime}}=0.3$, $T_{2,K[K^{\prime}]}$ $=0.1$,
$T_{3,K[K^{\prime}]}=0.3$, $P_{2[3],K[K^{\prime}]}$ $=0.4$, $\varphi
_{2(3),K[K^{\prime}]}^{R}=\pi$, $\varphi_{2[3],K[K^{\prime}]}^{T}=0$,
$\varphi_{1(4)}^{R}=\pi/2$, $\Delta\varphi_{2(3),K[K^{\prime}]}^{R\{T\}}=0$
and $\delta_{23}=\pi$. We have indicated the position of the resonances
$\mathcal{A}_{1(4)}^{m}$ and $\mathcal{A}_{3(5)}^{m}$ with red and blue dashed
lines, respectively.}%
\label{GraySetupA}%
\end{figure}

We now consider setup (a), which has been frequently used in the
MDI regime, for studying the spin accumulation
effect\cite{Johnson,Jedema,Jedema2,Zaffalon,Lou,Tombros2}. We first assume
that the $K$ and $K^{\prime}$ channels are coupled identically to the leads.
This case is illustrated by Fig. \ref{GraySetupAbis}, which shows the
variations of $G^{P}$ (top left panel), $V^{P}$ (top right panel), $MG$
(bottom left panel) and $MV$ (bottom right panel) versus $\delta_{12}$
(horizontal axes) and $\delta_{34}$ (vertical axes). One can first notice that
all these signals present strong variations with $\delta_{12}$ and
$\delta_{34}$, due to quantum interferences occurring inside the CNT. In Fig.
\ref{GraySetupAbis}, $G^{P}(\delta_{12})$ presents peaks which correspond to
the resonances $\mathcal{A}_{1}^{m}$ (see e.g. red dashed line), because we
consider a case where $T_{2m}$ is weak (these peaks also correspond
accidentally to the resonances $\mathcal{A}_{4}^{m}$, which are much broader).
A more remarkable result is that $G^{P}(\delta_{34})$ presents antiresonances
which correspond to $\mathcal{A}_{3}^{m}$ and $\mathcal{A}_{5}^{m}$ (see e.g.
blue dashed line). This is a signature of the strongly non-local nature of
current transport in this circuit: the electric signal measured in a given
section of the CNT can be sensitive to resonances occurring in other sections
of the CNT. We note that in Fig.~\ref{GraySetupAbis}, $\left\vert
V^{P}\right\vert $ presents the same type of variations as $G^{P}$ with
$\delta_{12}$ and $\delta_{34}$. In the general case, the resonances or
antiresonances shown by the electric signals will not necessarily correspond
to those defined in Fig. \ref{Reson4term}, due to the strong coupling between
these different types of resonances. Importantly, we find that the $MG$ signal
can be finite, contrarily to what happens in the MDI limit. Indeed, in Fig.
\ref{GraySetupAbis}, $MG$ can exceed $8\%$. We note that in Fig.
\ref{GraySetupAbis}, $MG$ presents minima approximately correlated with the
maxima of $G^{P}$ in the $\delta_{12}$ direction, and with the minima of
$G^{P}$ in the $\delta_{34}$ direction. In the case of a $S^{m}$ matrix
independent from $m$, one finds $V^{c}=0$ (see section \ref{ScatModel}). By
continuity, since we have used in Fig. \ref{GraySetupAbis} relatively low
values for $P_{2[3],K[K^{\prime}]}$, no SDIPS and symmetric $K$ and
$K^{\prime}$ channels, we find $\left\vert V^{P}\right\vert \ll V_{b}$. More
precisely, a lowest order development with respect to $P_{2}$ and $P_{3}$
yields $V^{P}\sim-V^{AP}\sim\lambda P_{2}P_{3}$, with $\lambda\ll1$ a function
of the different system parameters. In these conditions, $MV$ presents the
same type of variations as $V^{P}$ (one has $MV\sim2V^{P}$). When the $K$ and
$K^{\prime}$ modes are strongly asymmetric, it is possible to obtain a strong
$\left\vert V^{P}/V_{b}\right\vert $ ratio for relatively low polarizations
$P_{2[3],p}$. This case is illustrated by Fig. \ref{GraySetupA}, where we have
used $T_{1[4],K}\neq T_{1[4],K^{\prime}}$ and $\varphi_{1(4)}^{R}\neq0$, so
that $\mathcal{A}_{1[4]}^{(K,\sigma)}\neq\mathcal{A}_{1[4]}^{(K^{\prime
},\sigma)}$ and $\mathcal{A}_{3[5]}^{(K,\sigma)}\neq\mathcal{A}_{3[5]}%
^{(K^{\prime},\sigma)}$. In this case, the variations shown by the different
electric signals are more complicated than previously. However, we find
$V^{P}\sim V^{AP}$, so that the amplitude of $MV$ remains comparable to that
of Fig. \ref{GraySetupAbis}. 

\begin{figure}[ptbh]
\includegraphics[width=1.\linewidth]{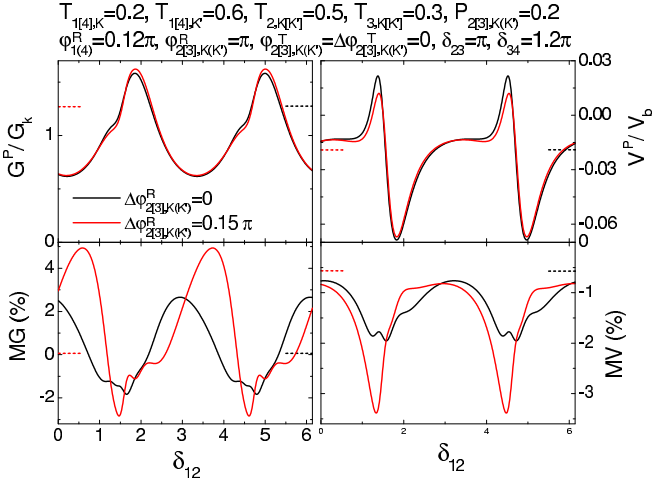}\caption{Signals $G^{P}$,
$V^{P}$, $MG$, and $MV$ as a function of $\delta_{12}$ for setup (a). We
consider a case with no SDIPS (black lines) and a case with a finite SDIPS
(red lines corresponding to $\Delta\varphi_{2(3),K[K^{\prime}]}^{R}=0.15\pi$).
We have used $T_{1[4],K}=0.2$, $T_{1[4],K^{\prime}}=0.6$, $T_{2,K[K^{\prime}%
]}$ $=0.5$, $T_{3,K[K^{\prime}]}=0.3$, $P_{2[3],K[K^{\prime}]}$ $=0.2$,
$\varphi_{1(4)}^{R}=0.12\pi$, $\varphi_{2(3),K[K^{\prime}]}^{R}=\pi$,
$\varphi_{2[3],K[K^{\prime}]}^{T}=\Delta\varphi_{2(3),K[K^{\prime}]}^{T}=0$,
$\delta_{23}=\pi$ and $\delta_{34}=0.12\pi$. The full lines correspond to the
CFC prediction (section \ref{ScatModel}), and the dotted lines to the IFC
prediction (section \ref{IFC}). The second does not depend on $\delta_{12}$.
In the IFC case, the MG signal is hardly visible in this figure because it is
of the order of 0.06\%.}%
\label{compSDIPS}%
\end{figure}

We now discuss the signs of the different signals. We have already seen above
that with the parameters of Fig. \ref{GraySetupAbis}, one has $V^{P}>0$ and
$V^{AP}<0$. In other conditions, it is possible to have $V^{P}<0$ and
$V^{AP}>0$, or $V^{P}$and $V^{AP}$ both positive, or both negative (not
shown). In the CFC model, the signs of $V^{P}$ and $V^{AP}$ are thus
independent, whereas MDI models usually give opposite signs for $V^{P}$ and
$V^{AP}$ (see e.g. Eq. (\ref{VCMDI}) of Appendix A and Refs.
\onlinecite{Johnson2, Takahashi}). Figure \ref{GraySetupA} illustrates that
there exists sets of parameters such that the non-local voltage $V^{P}$
changes sign while sweeping $\delta_{12}$ or $\delta_{34}$ (this result is
also true for $\delta_{23}$) \cite{SignChange}. It is also possible to find
sets of parameters such that $MV$ (not shown) and $MG$ (see Fig.
\ref{compSDIPS}, bottom left panel, full lines) change sign with $\delta_{12}%
$, $\delta_{23}$ or $\delta_{34}$.

We now briefly discuss the effects of the contacts polarizations. One can
generally increase the amplitude of the magnetic signals by increasing
$P_{j,p}$ (not shown), $\left\vert \Delta\varphi_{j,p}^{R}\right\vert $ (see
Fig. \ref{compSDIPS}, red full lines) and $\left\vert \Delta\varphi_{j,p}%
^{T}\right\vert $ (not shown). A strong SDIPS can split the resonances or
antiresonances of the electric signals (not shown), like already found in the
two-terminals $F$/CNT/$F$ case\cite{Cottet06a}. Interestingly, in the case of
a two-terminals $F$/CNT/$F$ device with a $K-K^{\prime}$ degeneracy and no
SDIPS (using $1=F$, $2=F$ and no leads $3$ and $4$),
Ref.~\onlinecite{Cottet06a} has found that the oscillations of $MG$ with
$\delta_{12}$ are symmetric, and a finite SDIPS is necessary to break this
symmetry. In contrast, in setup (a), the oscillations of $MG(\delta_{12})$ can
be asymmetric in spite of the $K-K^{\prime}$ degeneracy and the absence of a
SDIPS (see Fig. \ref{compSDIPS}, bottom left panel).

\subsection{Behavior of setup (b)}

\begin{figure}[ptbh]
\includegraphics[width=1.\linewidth]{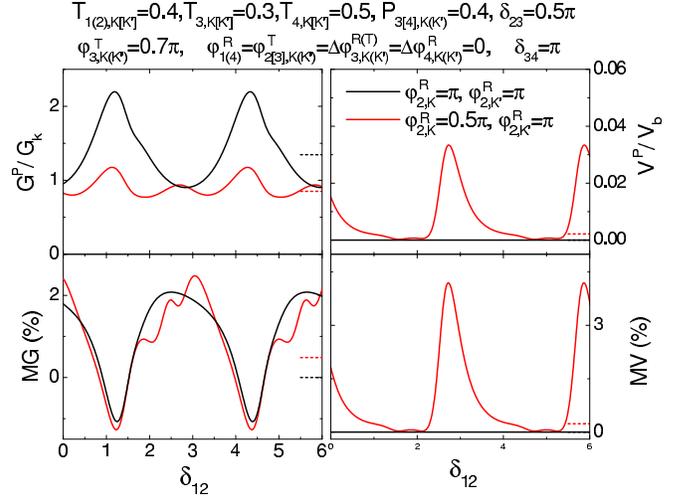}\caption{Signals $G^{P}$,
$V^{P}$, $MG$, and $MV$ as a function of $\delta_{12}$, for setup (b). The
black (red) lines correspond to the cases of $K$ and $K^{\prime}$ channels
coupled identically (differently) to the contacts, i.e. $\varphi_{2,K}%
^{R}=\varphi_{2,K^{\prime}}^{R}=\pi$ ($\varphi_{2,K}^{R}=1.2\pi$,
$\varphi_{2,K^{\prime}}^{R}=0.6\pi$). We have used $T_{1[2],K(K^{\prime}%
)}=0.4$, $T_{3,K[K^{\prime}]}=0.3$, $T_{4,K[K^{\prime}]}=0.5$,
$P_{3[4],K[K^{\prime}]}$ $=0.4$, $\varphi_{1(4)}^{R}=\varphi_{2(3),K[K^{\prime
}]}^{T}=\Delta\varphi_{3,K[K^{\prime}]}^{R(T)}=\Delta\varphi_{4,K[K^{\prime}%
]}^{R}=0$, $\delta_{23}=\pi/2$ and $\delta_{34}=\pi$. The full lines
correspond to the CFC prediction and the dotted lines to the IFC prediction.
In the black case, the $MG$ signal vanishes in the IFC limit, and the $V^{P}$
and $MV$ signals vanish in both the IFC and CFC limits. The MG signal of the
red case is hardly visible in the IFC limit because it is of the order of
0.03\%.}%
\label{setupB}%
\end{figure}

In setup (b), the types of resonances or antiresonances shown by the electric
signals depend again on the value of the coupling between the different CNT
sections. We will only highlight the most interesting specificities of setup
(b), because it has many common properties with setup (a). Fig. \ref{setupB}
shows, with black [red] full lines, examples of $G^{P}(\delta_{12})$,
$MG(\delta_{12})$, $V^{P}(\delta_{12})$, and $MV(\delta_{12})$ curves, for
symmetric [asymmetric] $K$ and $K^{\prime}$ channels . Strikingly, in both
cases, the magnetoconductance $MG$ between the $N$ leads 1 and 2 can be
finite, although the two $F$ leads are located outside the classical current
path. This is in strong contrast with the MDI limit. From Eqs. (\ref{Vnl4pat}%
-\ref{S32quat}), the voltage difference $V^{c}$ vanishes if the scattering
properties of contacts $1$ or $2$ are independent from the transverse index
$p$, regardless of the scattering properties of contacts $3$ and
$4$\cite{check}. This leads to the paradoxical situation where a magnetic
signal can be measured between the two $N$ leads but not between the two $F$
leads (see black full lines in Fig. \ref{setupB}).\ By continuity, when the
$K-K^{\prime}$ asymmetry is not large at contacts 1 and 2, the amplitude of
the signals $V^{P}$ and $MV$ measured between contacts $3$ and $4$ will remain
very small. It is possible to obtain stronger amplitudes for $V^{c}$ and $MV$
in the opposite limit of strongly asymmetric $K$ and $K^{\prime}$ channels
(see red full lines in Fig. \ref{setupB}, for which we have used
$\varphi_{2,K}\neq\varphi_{2,K^{\prime}}$). With setup (b), is thus also
possible to obtain magnetic signals in both $G^{c}$ and $V^{c}$, whereas $MG$
and $MV$ would vanish in the $MDI$ limit.

\subsection{Comparison with the MDI limit}

In this section, we summarize the most striking differences between the
coherent four-channels (CFC) model of section \ref{ScatModel} and the MDI
model of section \ref{MDI}. For setup (b), the CFC model allows $V^{P}\neq0$
and $MV\neq0$ whereas one finds $V^{P}=0$ and $MV=0$ with the MDI model.
Another remarkable result is that for both setups (a) and (b), the CFC model
gives $G^{P}\neq G^{AP}$ whereas the MDI model imposes $G^{P}=G^{AP}$. The
table in Fig. \ref{Circuits} summarizes these results.

\section{Incoherent four-channels limit\label{IFC}}

In order to determine whether the specific spin-dependent behavior of the CFC
model is due to coherence or to the low number of channels, it is interesting
to consider the incoherent four-channels (IFC) limit. If the phase relaxation
length of the CNT is much shorter than the distance between the different
contacts, the global transmission and reflection probabilities of setups (a)
and (b) can be calculated by composing the scattering probabilities of the
different contacts instead of the scattering amplitudes\cite{Datta}. We have
checked that this leads to replacing the scattering probabilities $\left\vert
S_{\alpha\beta}^{m}(\{r_{jm},t_{jm},v_{jm},u_{jm},\delta_{ij}\})\right\vert
^{2}$ occurring in Eqs.(\ref{I}) and (\ref{Gdef}) by $S_{\alpha\beta}%
^{m}(\{\left\vert r_{jm}\right\vert ^{2},\left\vert t_{jm}\right\vert
^{2},\left\vert v_{jm}\right\vert ^{2},\left\vert u_{jm}\right\vert ^{2}%
,0\})$. Importantly, this description remains intrinsically quantum since the
channel quantization is taken into account. In Figs. \ref{compSDIPS} and
\ref{setupB}, we show with black and red dotted lines the IFC values
corresponding to the different CFC curves. We find that $G^{c}$, $V^{c}$, $MG$
and $MV$ do not depend anymore on the phases $\delta_{ij}$. However,
$G^{P}\neq G^{AP}$ is still possible for setups (a) and (b). More precisely,
we have checked analytically that using identical $K$ and $K^{\prime}$ modes
leads to\cite{check} $G^{P}=G^{AP}$, and we have checked numerically that
$G^{P}\neq G^{AP}$ occurs in case of a $K$/$K^{\prime}$ asymmetry at one of
the four contacts for setup (a), and at contacts 1 or 2 for setup (b). We can
also obtain $V^{P}\neq0$ and $MV\neq0$ for setup (b) [and, more trivially, for
setup (a)], with the same symmetry restrictions as for the CFC case (see table
in Fig. \ref{Circuits}). Therefore, having $V^{P}\neq0$ and $MV\neq0$ for
setup (b), and $MG\neq0$ for setups (a) and (b) is not a specificity the
coherent case: using a very small number of transport channels already allows
these properties. It is nevertheless important to notice that the values of
$MG$ and $MV$ are strongly enhanced in the CFC case, due to resonance effects.
Moreover, in the IFC case, the circuit is insensitive to the SDIPS, whereas in
the coherent case, the SDIPS furthermore increases the amplitude of $MV$ and
$MG$\cite{IFCnoSDIPS}. At last, the coherent case presents the interest of
allowing strong variations of the electric signals with the gate-controlled
phases $\delta_{12}$, $\delta_{23}$ and $\delta_{34}$.

\section{Discussion on first experiments}

Reference~\onlinecite{Tombros} reports on a Single Wall Carbon Nanotube (SWNT)
circuit biased like in Fig. \ref{Circuits}, but with four ferromagnetic leads.
A hysteretic $V^{c}$ has been measured by flipping sequentially the
magnetizations of the two inner contacts. However, no conclusion can be drawn
from this experiment, due to the lack of information on the conduction regime
followed by the device. Reference \onlinecite{Gunnar} reports on $V^{c}$
measurements for a setup (a) made with a SWNT. The authors of this Ref. have
observed a finite $V^{c}$ which oscillates around zero while the back gate
voltage of the sample is swept. This suggests that this experiment was in the
coherent regime. However, the amplitude of $V^{c}$ was very low ($\max
(\left\vert V^{c}/V_{b}\right\vert )\sim0.01$), which indicates, in the
framework of the scattering model, that the $K$ and $K^{\prime}$ modes were
very close and the spin polarization of the contacts scattering properties
very weak. It is therefore not surprising that these authors did not obtain a
measurable $MV$ signal. Although setup (a) seems very popular in the
nanospintronics community for historical reasons \cite{Johnson}, we have shown
above that setup (b) also highly deserves an experimental effort, as well as
$MG$ measurements in general.

In this article, we have chosen to focus on the case of double mode quantum
wires because this is adapted for describing CNT based devices, which are
presently among the most advanced nanospintronics devices. However,
technological progress might offer the opportunity to observe the effects
depicted in this article in other types of nanowires, like e.g. semiconducting
nanowires. Indeed, quantum interferences have already been observed in
\textrm{Si}\cite{Tilke} and \textrm{InAs} quantum wires\cite{Doh}, and
spin-injection has already been demonstrated in \textrm{Si} layers
\cite{Jonker} and \textrm{InAs} quantum dots\cite{Hamaya}. One major
difficulty may consist in reaching the few modes and fully
ballistic\cite{Zhou} regime with these devices.

\section{Conclusion}

In this work, we have studied theoretically various circuits consisting of a
carbon nanotube with two transverse modes, contacted to two normal metal leads
and two ferromagnetic leads. Two contacts are used as source and drain to
define a local conductance, and the two other contacts are left floating, to
define a non-local voltage outside the classical current path. When the
magnetizations of the two ferromagnetic contacts are changed from a parallel
to an antiparallel configuration, we predict, in the local conductance and the
non-local voltage, magnetic signals which are specific to the case of a system
with a low number of channels. In particular, we propose an arrangement of the
normal and ferromagnetic leads [setup (b)] which would give no magnetic
response in the multichannel diffusive incoherent (MDI) limit, but which
allows magnetic responses in both the local conductance and the non-local
voltage in the two-modes regime. The more traditional arrangement [setup (a)]
used for the study of the MDI limit also shows a qualitatively new behavior,
i.e. a magnetic response in the local conductance. These specific magnetic
behaviors are strongly reinforced in the coherent case, due to resonance
effects occurring inside the nanotube, and also, possibly, due to the
Spin-Dependence of Interfacial Phase Shifts. Our calculations pave the way to
new experiments on non-local spin-transport in low-dimensional conductors.

We acknowledge discussions with H. U. Baranger and G. E. W. Bauer. This work
was financially supported by the ANR-05-NANO-055 contract, the European Union
contract FP6-IST-021285-2 and the C'Nano Ile de France contract SPINMOL.

\section{Appendix A: Discussion of the MDI limit with a resistors network}

\begin{figure}[ptbh]
\includegraphics[width=0.6\linewidth]{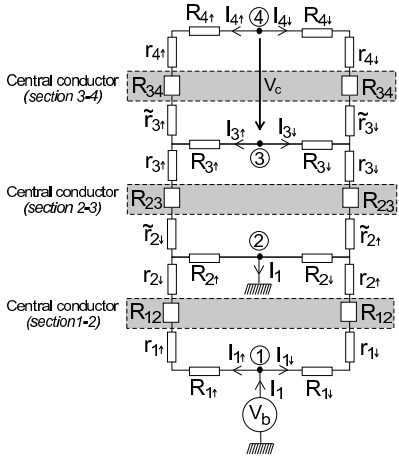}\caption{Resistors
network used to describe the behavior of a one dimensional Central Conductor
(CC) connected to four leads $j\in\{1,2,3,4\}$ in the MDI limit. The contact
between lead $j$ and the CC is represented by the resistors $R_{j}^{\sigma}$,
$r_{j}^{\sigma}$, and $\widetilde{r}_{j}^{\sigma}$. The section $j-k$ of the
CC is modeled with the two resistors $R_{jk}$. When lead $j$ is not
ferromagnetic, one must use $R_{j}^{\uparrow}=R_{j}^{\downarrow}$,
$r_{j}^{\uparrow}=r_{j}^{\downarrow}$, and $\widetilde{r}_{j}^{\uparrow
}=\widetilde{r}_{j}^{\downarrow}$}%
\label{Resistances.eps}%
\end{figure}

In this appendix, we discuss the MDI regime with an elementary but insightful
resistors network model. When the electronic mean free path is much shorter
than the spin-flip length, it is possible to define a spin-dependent
electrochemical potential which obeys a local spin-dependent Ohm's
law\cite{Zutic}. Thus, neglecting spin-flip scattering inside the CC, one can
use the effective resistor network of Fig. \ref{Resistances.eps} to describe
the behaviors of setups (a) and (b) in the MDI limit\cite{Valet,Tombros}. For
completeness, we allow the four leads $j\in\{1,2,3,4\}$ to be ferromagnetic,
with colinear magnetizations. The left (right) part of the resistors network
corresponds to the up (down) spin channels. Due to intra-lead spin-flip scattering, electrons are in local
equilibrium in lead $j$. This equilibration is described by the electrical
connection of $\uparrow$ and $\downarrow$ channels at node $j$, which has an
electric potential $V_{j}^{c}$. The section $j-k$ of the CC is modeled with
the two resistors $R_{jk}$. The contact between lead $j$ and the CC is
represented by the resistors $R_{j}^{\sigma}$, $r_{j}^{\sigma}$, and
$\widetilde{r}_{j}^{\sigma}$. When lead $j$ is not ferromagnetic, one must use
$R_{j}^{\uparrow}=R_{j}^{\downarrow}$, $r_{j}^{\uparrow}=r_{j}^{\downarrow}$,
and $\widetilde{r}_{j}^{\uparrow}=\widetilde{r}_{j}^{\downarrow}$. The current
flowing from lead $1$ to lead $2$ is $I_{1}^{c}=I_{1\uparrow}^{c}%
+I_{1\downarrow}^{c}$. Since leads $3$ and $4$ are floating, they supply the
CC with spin currents which are perfectly equilibrated, i.e. $I_{j\uparrow
}=-I_{j\downarrow}$ for $\alpha\in\{3,4\}$. We find
\begin{equation}
G^{c}=\mathcal{H}V_{b}/\mathcal{D}\label{GCMDI}%
\end{equation}
and%
\begin{equation}
V^{c}=\left(  A_{34}^{\uparrow}-A_{34}^{\downarrow}\right)  \left(
B_{12}^{\uparrow}-B_{12}^{\downarrow}\right)  V_{b}/\mathcal{R}_{23}%
\mathcal{H}\label{VCMDI}%
\end{equation}
with
\begin{equation}
\mathcal{R}_{23}=\frac{R_{3}^{\uparrow}+R_{3}^{\downarrow}}{\sum
\limits_{\sigma}\left(  \mathcal{R}_{34}^{\sigma}-R_{3}^{\sigma}\right)
}+\sum\limits_{\sigma}\left(  \widetilde{r}_{2}^{\sigma}+r_{3}^{\sigma}%
+R_{23}\right)
\end{equation}%
\begin{equation}
\mathcal{H}=\sum\limits_{\sigma}\left(  \mathcal{R}_{12}^{\sigma}%
+\alpha_{\sigma,\sigma}+\alpha_{\sigma,-\sigma}\right)
\end{equation}%
\begin{equation}
\mathcal{D}=\left(  \mathcal{R}_{12}^{\uparrow}+\alpha_{\uparrow,\uparrow
}\right)  \left(  \mathcal{R}_{12}^{\downarrow}+\alpha_{\downarrow,\downarrow
}\right)  -\alpha_{\uparrow,\downarrow}\alpha_{\downarrow,\uparrow}%
\end{equation}%
\begin{equation}
2A_{34}^{\sigma}=\mathcal{R}_{34}^{\sigma}\left[  (R_{3}^{\uparrow}%
+R_{3}^{\downarrow})/(\mathcal{R}_{34}^{\uparrow}+\mathcal{R}_{34}%
^{\downarrow})\right]  -R_{3}^{\sigma}%
\end{equation}%
\begin{equation}
B_{12}^{\sigma}=\left(  \mathcal{R}_{12}^{\sigma}+\alpha_{\sigma,\sigma
}+\alpha_{-\sigma,\sigma}\right)  \left(  \mathcal{R}_{12}^{-\sigma}%
-R_{2}^{-\sigma}\right)
\end{equation}%
\begin{equation}
\alpha_{\sigma,\sigma^{\prime}}=R_{2}^{\sigma}(\mathcal{R}_{12}^{\sigma
^{\prime}}-R_{2}^{\sigma^{\prime}})/\mathcal{R}_{23}%
\end{equation}
and%
\begin{equation}
\mathcal{R}_{jk}^{\sigma}=R_{j}^{\sigma}+\widetilde{r}_{j}^{\sigma}%
+R_{k}^{\sigma}+r_{k}^{\sigma}+R_{jk}%
\end{equation}
for $(j,k)\in\{1,2,3,4\}^{2}$. The value of $\mathcal{H}$ is independent from
the contacts magnetic configuration, but $\mathcal{D}$ depends on the relative
configuration of leads $1$ and $2$, so that $G^{P}\neq G^{AP}$ is possible
provided leads $1$ and $2$ are ferromagnetic. In contrast, the value of
$G^{c}$ is independent from the magnetization directions of leads $3$ or $4$
because, due to $I_{3(4)\uparrow}=-I_{3(4)\downarrow}$, the resistors
$R_{3}^{\sigma}$, $r_{3}^{\sigma}$, $\widetilde{r}_{3}^{\sigma}$,
$R_{4}^{\sigma}$ and $r_{4}^{\sigma}$ of Fig. \ref{Resistances.eps} are
connected in series with $R_{3}^{-\sigma}$, $r_{3}^{-\sigma}$, $\widetilde
{r}_{3}^{-\sigma}$, $R_{4}^{-\sigma}$ and $r_{4}^{-\sigma}$ respectively. We
conclude that for setups (a) and (b), one has $G^{P}=G^{AP}$ in the MDI limit.
From Eq. (\ref{VCMDI}), having $V^{P}\neq0$ requires that at least one of the
biased leads $1$ or $2$ is ferromagnetic (for the generation of a
spin-accumulation), and at least one of the floating leads $3$ or $4$ is
ferromagnetic (for the detection of this spin-accumulation). These conditions
are fulfilled for setup (a), but not for setup (b). Importantly, these results
will not be modified if a moderate intra-CC spin-flip scattering or the finite
width of the contacts is taken into account, because both features can be
modelled with a distributed array of resistors connecting the two spin
branches, which will not change the spin symmetry of the model of Fig.
\ref{Resistances.eps}.


\begin{thebibliography}{99}
\bibitem {simpl}From Eq. (\ref{S11quat}), the phase of $t_{1m}t_{1m}^{\prime
}/r_{1m}^{\prime}$ can also affect the signals $G^{c}$ and $V^{c}$. However,
since the scattering matrix of contact 1 is unitary, one finds $t_{1m}%
t_{1m}^{\prime}/r_{1m}^{\prime}=-r_{1m}|t_{1m}/r_{1m}|^{2}$. It is thus
sufficient to use $\arg(r_{1m})$ as a parameter.
\bibitem {param}The complete parametrization of the scattering matrices of
contacts $j\in\{2,3\}$ requires extra parameters $\varphi_{j,p}^{U}$, and
$\Delta\varphi_{j,p}^{U}$ such that $\arg(u_{j(p,\sigma)})=\varphi_{j,p}%
^{U}+\sigma\Delta\varphi_{j,p}^{U}/2$ and $r_{jm}^{\prime\prime}%
=-(1-2|u_{jm}|^{2})^{1/2}e^{i\arg[(u_{jm})^{2}/(A_{j}^{m}+B_{j}^{m})]}$.
However, one can check that, with the assumptions made in part IV.B, the
signals $G^{c}$ and $V^{c}$ are independent from $\varphi_{j,p}^{U}$ and
$\Delta\varphi_{j,p}^{U}$ for both setups (a) and (b).
\bibitem {annul}For completeness, we point out that the condition on $\phi
_{j}^{m}$ can be relaxed if $t_{j}^{m}=0$.
\bibitem {sym}For $j\in\{2,3\}$, one has $\left\vert u_{jm}\right\vert
^{2}\leq1/2$ because we assume that an electron arriving from contact $j$
enters the top and bottom portions of the CNT with the same probabilities.
\bibitem {SignChange}We find that the sign changes of $V^{P}$ can already
occur in the spin-degenerate case. Accordingly, a change of sign in $V^{P}$
has already been observed for a CNT with four normal metal leads, by sweeping
the CNT back-gate voltage\cite{Makarovski}.
\bibitem {check}We have checked analytically that this property is true even
when the contacts 2 and 3 are spatially asymmetric i.e. $u_{(2)3m}\neq
v_{(2)3m}$.
\bibitem {IFCnoSDIPS}In the IFC case, $MV$ and $MG$ are sensitive to the
$\Delta\varphi_{2(3),p}^{R[T]}$ parameters through the reflection
probabilities of contacts $2$ and $3$, but not through the scattering phases.
\end{thebibliography}
\end{document}